\documentclass[%
 aip,
 amsmath,amssymb,
 reprint,%
]{revtex4-1}

\usepackage{graphicx}
\usepackage{dcolumn}
\usepackage{bm}

\usepackage[utf8]{inputenc}
\usepackage[T1]{fontenc}
\usepackage{mathptmx}

\usepackage{amssymb}
\usepackage{subfig}
\usepackage{tikz}

\begin{document}

\preprint{AIP/123-QED}

\title[Machine Learning with bond information for local structure optimizations in surface science]{Machine Learning with bond information for local structure optimizations in surface science}

\author{Estefan\'ia Garijo del R\'io}
\affiliation{Department of Physics, Technical University of Denmark, Kgs. Lyngby, Denmark}

\author{Sami  Kaappa}
\affiliation{Department of Physics, Technical University of Denmark, Kgs. Lyngby, Denmark}

\author{Jos\'e A. Garrido Torres}
\affiliation{SUNCAT Center for Interface Science and Catalysis, Department of Chemical Engineering,\\ Stanford University, Stanford, CA 94305, USA}
\affiliation{SLAC National Accelerator Laboratory, 2575 Sand Hill Road, CA 94025, USA}
\affiliation{Columbia Electrochemical Energy Center, Department of Chemical Engineering,\\ Columbia University, New York, NY 10027, USA}

\author{Thomas Bligaard}
\affiliation{SUNCAT Center for Interface Science and Catalysis, Department of Chemical Engineering,\\ Stanford University, Stanford, CA 94305, USA}
\affiliation{Department of Energy Conversion and Storage, Technical University of Denmark, Kgs. Lyngby, Denmark}

\author{Karsten Wedel Jacobsen}
\email{kwj@fysik.dtu.dk}
\affiliation{Department of Physics, Technical University of Denmark, Kgs. Lyngby, Denmark}

\date{\today}

\begin{abstract}
  Local optimization of 
  adsorption systems inherently involves different scales: within the substrate, within the molecule, and between molecule and substrate. In this work, we show how the explicit modeling of the different character of the bonds in these systems improves the performance of 
  machine learning methods for optimization. We introduce an anisotropic kernel in the Gaussian process regression framework that guides the search for the local minimum, and we show its overall good performance across different types of atomic systems. The method shows a speed-up of up to a factor two compared with the fastest standard optimization methods on adsorption systems. Additionally, we show that a limited memory approach is not only beneficial in terms of overall computational resources, but can  result in a further reduction of energy and force calculations
\end{abstract}

\maketitle


\section{Introduction}

One of the most common tasks in computational heterogeneous catalysis is finding local minima in a potential energy surface (PES). Such equilibrium atomic configurations are of great interest, since they are often the first step from which more complicated studies of reaction rates are carried out. A number of well-established methods exist for this task \cite{nocedal2006numerical,FIRE,lindh1995precon,burger2010quasi}, which rely on iteratively computing energy-force pairs for a set of atomic configurations. A common and successful choice for the computation of energies and forces is density functional theory (DFT) \cite{Hohenberg:1964fz,Kohn:1965js}. Even though this approach carries a good trade-off between computational cost and accuracy, the structure determination can be very computationally demanding if the optimization method requires many energy-force evaluations.

Recently, the field of efficient local optimization of atomic structures has attracted considerable attention. An interesting approach is that of \emph{preconditioning} \cite{2016CsanyiPrecon,mones2018preconditioners,Makri2019precon}: in atomic systems where bonds of very different stiffness are present, changes of certain atomic positions produce much more rapid changes in energy than others, and this can result on a slowdown of traditional methods. If the differences in stiffness are very large, the forces may not point in the direction towards the minimum, in what is known as a poorly scaled optimization problem \cite{nocedal2006numerical}. Preconditioning then consists in finding a linear transformation of the problem that will lead to a Hessian with a better condition number, which corrects for the difference in stiffness in the PES landscape and results in a better guidance of the search. For atomic systems, preconditioners based on the adjacency matrix of atoms and their interatomic distances \cite{2016CsanyiPrecon} or on the Hessian of semi-empirical potentials \cite{mones2018preconditioners} have shown a significant reduction of the number of steps necessary to relax atomic structures, as well as to guide transition state searches \cite{Makri2019precon}. 

The use of machine learning techniques to build surrogate models of the PES that are then used to guide the search of critical points has also recently attracted increasing attention. Successful examples of methods that have achieved a significant reduction of the number of electronic structure evaluations are abundant for local optimization \cite{denzelGPR2018,Schmitz2018internalnograd,GPMin,Hauser2020Internal,shapeev:ActiveLearning,shapeev:alloys}, as well as transition state search \cite{Peterson2016saddle,koinstinen2017,koistinen2019nudged,denzel2018transition,denzel2020hessian,GarridoPRL2019} and global optimization of atomic structures \cite{Bjrok2018GPR-EA,Deringer2018boron,BO-rrss-2018,todorovic2019bayesian,Malthe2020efficient,Mortensen2020reinforced,fang2020efficient}.

In particular, Gaussian process regression (GPR) \cite{williams2006gaussian} has proved itself a particularly successful technique to build the surrogate PES that guides the critical point search, since it has the ability to generalize given just a few training points.

The computation of forces comes with little additional computational overhead to the energy computation
in DFT, and training on both energies and forces has become a well-established technique in the field \cite{denzelGPR2018,GPMin,Hauser2020Internal,koinstinen2017,koistinen2019nudged,denzel2018transition,GarridoPRL2019}. Along these lines, there has been a recent attempt to also incorporate higher derivatives \cite{denzel2020hessian}.
A recent study by Christensen and von Lilienfeld \cite{christensen2020gradients} has confirmed that the inclusion of forces along with energies of the configurations as targets in the training set results in a significant increase in the precision of the surrogate model of the PES of a single atomic system.

A less well established choice is that of the correlation model between two atomic structures or the kernel in the case of GPR. After the initial success in the use of stationary covariance functions of Cartesian coordinates (squared exponential and Mat\'ern covariance functions) \cite{denzelGPR2018,GPMin,koinstinen2017,denzel2018transition,GarridoPRL2019}, there have been some studies attempting to extend these covariance functions in order to further reduce the number of DFT calculations needed to find the critical point. Koistinen \emph{et al.} \cite{koistinen2019nudged} have proposed a non-stationary kernel based on the difference between the inverse of interatomic distances in each configuration for each pair of atoms. Meyer and Hauser \cite{Hauser2020Internal} have instead proposed the use of the squared exponential and Mat\'ern kernels in internal coordinates, instead of Cartesian. Both approaches have lead to a further reduction of steps. We note that outside the subfield of gradient-based GPR modeling for PES critical point identification, both  internal coordinates \cite{Schmitz2018internalnograd, Deringer2017machine, fang2020efficient} and fingerprints \cite{Malthe2020efficient,christensen2020gradients,Chmiela2016FF,GAP,Bartok2013representing,lilienfeld2012coulomb,lilienfeld2015bagofbonds,jager2018machine} have been used to incorporate knowledge of the PES topology into the covariance function of kernel methods.

In this paper we introduce a preconditioning scheme of the usual squared exponential kernel in Cartesian coordinates. The resulting expression for the kernel we propose can be reinterpreted in terms of chemical bonds and covalent radii, making it easy for the method to account for differences in the stiffness of each interaction and easy for the user to interpret the results. In this way, the method relies on a model of bond stiffness that can be provided by the user, but we prove that an educated guess can work even better if the method is allowed to self-update and find the bond constants itself. In addition, the structure of the kernel naturally incorporates the translation invariance of the PES. 

We have incorporated the new surrogate model into a machine learning optimization method that we have named BondMin and we have tested its performance in local relaxation problems with DFT. For this method, we have obtained speed-ups of up to a factor 2 for problems that involve molecules on surfaces as compared to the quasi-Newton method BFGSLineSearch, while retaining the good performance of the not preconditioned squared exponential kernel on general atomic systems.

\section{Methods}

\subsection{Gaussian process regression} 

Let $\mathbf{r}_i$ stand for the position vector of the $i$-th atom. For each atomic configuration $\mathbf{x} = \left( \mathbf{r}_1, \mathbf{r}_2, \dots, \mathbf{r}_{N_\text{atoms}}\right)$, we describe the surrogate potential energy surface (sPES) $E(\mathbf{x})$ and the associated force field $\mathbf{f}(\mathbf{x})$ using Gaussian process regression (GPR) \cite{williams2006gaussian}:
\begin{equation}
    (E(\mathbf{x}),-\mathbf{f}(\mathbf{x})) \sim \mathcal{N}\left(\mathbf{m}(\mathbf{x}), K(\mathbf{x}, \mathbf{x}^\prime) \right),
\end{equation}
where $\mathbf{m}(\mathbf{x})$ is the prior mean for each variable and $K(\mathbf{x}, \mathbf{x}^\prime)$ is the prior covariance matrix. This matrix can be written in terms of the kernel function $k(\mathbf{x}, \mathbf{x}^\prime )$ as \cite{poloczek2017gradients}:
\begin{equation}
    K(\mathbf{x}, \mathbf{x}^\prime) = \left( \begin{matrix}
    k(\mathbf{x}, \mathbf{x}^\prime) & \left(\nabla_{\mathbf{x}^\prime} k(\mathbf{x}, \mathbf{x}^\prime)\right)^T\\
    \nabla_{\mathbf{x}} k(\mathbf{x}, \mathbf{x}^\prime) &  \nabla_{\mathbf{x}}\left( \nabla_{\mathbf{x}^\prime}k(\mathbf{x}, \mathbf{x}^\prime)\right)^T
    \end{matrix}\right).
\end{equation}

 The GPR is trained on density functional 
 theory (DFT) energies 
 $\{E^{(i)}\}{}_{i=1}^N$ and forces 
 $\{\mathbf{f}^{(i)}\}{}_{i=1}^N$ 
 corresponding to a set of $N$ atomic configurations 
 $\{\mathbf{x}^{(i)}\}{}_{i=1}^N$.
 We arrange the inputs into the 
 $3N_\text{atoms} \times N$
 design matrix $X$ and the targets
 $\{(E^{(i)}, -\mathbf{f}^{(i)})\}{}_{i=1}^N$
 into the $(3N_\text{atoms} +1) \times N$ matrix $Y$. 
 By denoting the Gram matrix as $K(X,X)$, which is given by the block matrices $(K(X,X,))_{ij} = K(\mathbf{x}^{(i)},\mathbf{x}^{(j)})$, and defining the matrix $K(\mathbf{x},X) = (K(\mathbf{x},\mathbf{x}^{(1)}),\dots, 
 K(\mathbf{x},\mathbf{x}^{(N)})  )$, the prediction can be written as:
 \begin{equation} \label{eq:prediction}
(E(\mathbf{x}),-\mathbf{f}(\mathbf{x})) = \mathbf{m}(\mathbf{x}) + K(\mathbf{x}, X) \; C_X^{-1} (Y-\mathbf{m}(X)), 
\end{equation}
where $C_X = K(X,X) + \Sigma$ is the regularized Gram matrix and the  diagonal matrix $\Sigma$ is the regularization. 

The GPR framework also includes an analytical expression for the marginal likelihood $p(Y\vert X)$ :
\begin{equation} \label{eq:mll}
\begin{split}
    \log p (Y\vert X) = & -\frac{1}{2} \left(Y-\mathbf{m}(X)\right)^T C_X^{-1} \left(Y-\mathbf{m}(X)\right) \\
                        & -\frac{1}{2} \log \vert C_X \vert + \mathfrak{N},
\end{split}
\end{equation}
which depends on a number of hyperparameters $\theta$ that parametrize the regularized kernel $C_X (X; \theta)$ and the prior $\mathbf{m}(X; \theta)$. The logarithm of the marginal likelihood can be maximized using a gradient-based optimizer to find the most likely hyperparameters given the inputs and the targets. $\mathfrak{N}$ stands for the normalization factor, that does not depend on $X$, $Y$ or $\theta$.

In this work we introduce a new kernel that uses the difference between the positions between every pair of atoms in the system to define a distance measure $d(\mathbf{x}, \mathbf{x}^\prime)$  between configurations:
\begin{equation} \label{eq:dist}
    d^2(\mathbf{x}, \mathbf{x}^\prime) = \frac{1}{N_\text{atoms}}\sum_{i,j}^{N_\text{atoms}} \frac{\left \Vert ( \mathbf{r}_i -\mathbf{r}_j) - ( \mathbf{r}_i^\prime -\mathbf{r}_j^\prime) \right\Vert^2}{\ell_{X_iX_j}^2},
\end{equation}
where $X_i$ stands for the atomic symbol of the $i$-th atom and $\mathbf{r}_i$ for its position. The scales $\ell_{X_i X_j}$ for each pair of atoms here have length dimension and have the role of re-scaling the weight of each interatomic distance according to the atomic type.

We note that equation \eqref{eq:dist} can be rewritten into matrix form as follows:
\begin{equation} \label{eq:dist_matrix}
    d^2(\mathbf{x}, \mathbf{x}^\prime) = \frac{1}{N_\text{atoms}} (\mathbf{x} -\mathbf{x}^\prime)^T G (\mathbf{x} -\mathbf{x}^\prime).
\end{equation}
It is easy to show that the metric matrix is given by $G = P^T \text{diag}(g,g,g) P$, where $P$ is the permutation matrix mapping $P \mathbf{x} =  (r_1^{(x)}, r_2^{(x)}, \dots, r_N^{(x)}, r_1^{(y)}, \dots, r_N^{(z)}) $ and $\text{diag}(g,g,g)$ is the diagonal block matrix composed by three copies of:
\begin{equation} \label{eq:g}
    g_{ij} = \begin{cases}
    \sum_{k\neq i} \ell^{-2}_{X_iX_k}  & \text{if } i=j,\\
    -\ell^{-2}_{X_iX_j} & \text{if } i\neq j.
    \end{cases}
\end{equation}

We note that matrix $g$ is the Laplacian matrix of a fully-connected undirected graph 
where the nodes are the atoms in the unit cell and the weights on the edges depend on the chemical species of the atoms that they connect as
$1/\ell_{X_iX_j}^2$.

This distance measure has then been incorporated into the usual squared exponential kernel, replacing the Euclidean distance between Cartesian coordinates:
\begin{equation} \label{eq:sqexp}
    k(\mathbf{x}, \mathbf{x}^\prime) = k_0^2 \exp (-d^2(\mathbf{x},\mathbf{x}^\prime)/2\ell^2)
\end{equation}
where $k_0$ and $\ell$ are hyperparameters: the prefactor of the kernel and the dimensionless global scale.

\begin{figure}[t]
\centering



	\includegraphics{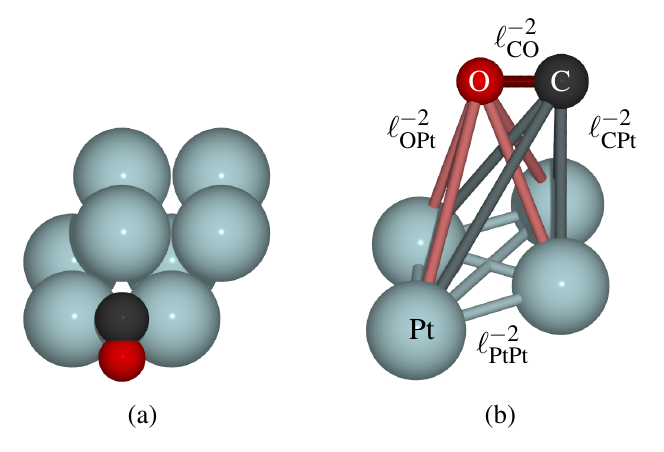}
\caption{ (a) Top view of the atomic structure of CO on a $2\times 2\times 2$ fcc (100) platinum slab. Only the atoms in the unit cell are shown.(b) Representation of the weighted fully-connected graph corresponding to the CO molecule and the platinum atoms in the top layer of the structure displayed in (a).}
\label{fig:diagram}
\end{figure}

One could define the vector $\mathbf{b}_{ij} = \mathbf{r}_i - \mathbf{r}_j$ in equation \eqref{eq:dist} as the bond vector defining the distance and the orientation of the bond between atoms $i$ and $j$ . In this picture, the distance between two configurations is then the weighted sum of Euclidean distances between all bonds, with $\ell_{X_i X_j}$ being the weight. However, note that in this conception every atom is bonded to every other atom in the atomic structure, so that the distance measure is not biased towards the initial structure. We note that the inclusion of the interatomic distance of every pair of atoms in the structure is frequently used in fingerprints 
(such as the Coulomb matrix \cite{lilienfeld2012coulomb} and other Coulomb-based definitions \cite{Bartok2013representing,HIMANEN2020Ds} or the bag of bonds \cite{lilienfeld2015bagofbonds}) and kernels \cite{Deringer2017machine, koistinen2019nudged} by the machine learning for materials and molecules community. We illustrate this concept in Figure \ref{fig:diagram}.

Equations \eqref{eq:dist_matrix} and \eqref{eq:sqexp} (and noting from equation \eqref{eq:dist} that the matrix $G$ is positive semi-definite for any value of the bond scales $\ell_{X_iX_j}$)  reveal that the kernel in terms of bonds is nothing but an anisotropic version of the stationary squared exponential kernel \cite{williams2006gaussian}. $G$ has three and only three zero eigenvalues, corresponding to translation along the three axis, making the method translationally invariant. Along the ideas in the work by Packwood \emph{et al.} \cite{2016CsanyiPrecon}, we note that $G$ can be factorized as $G=Q^TQ$ and that by defining the fingerprint $\mathbf{u} (\mathbf{x}) = Q \mathbf{x}$
one can regard $G$ as a preconditioner since the energy becomes less anisotropic  and hence better conditioned function in fingerprint space than in coordinate space.

For the particular case of a unary material, there is only one bond scale, $\ell_{XX}$. It can be shown that matrix $g$ as defined in equation \eqref{eq:g} has all eigenvalues equal to $N_\text{atoms}$, except for the one associated to translation symmetry (for example, by realizing $g$ becomes a circulant matrix 
for unaries). Then, if $\mathbf{x}$ and $\mathbf{x}^\prime$ do not differ in a translation, the distance in equation \eqref{eq:dist_matrix} becomes 
$d^2(\mathbf{x}, \mathbf{x}^\prime) = \Vert \mathbf{x} -\mathbf{x}^\prime \Vert/\ell_{XX}^2$
and the kernel in \eqref{eq:sqexp} becomes the isotropic squared exponential kernel 
$k(\mathbf{x}, \mathbf{x}^\prime)= k_0^2\exp{\Vert\mathbf{x}-\mathbf{x}^\prime\Vert^2/2\ell^2\ell_{XX}^2}$.
Consequently, if the sampling method used to generate the training set does not generate global translations of the atomic structure (i.e. the optimizer does not translate the system), an active learning method using it would behave as its isomorphic squared exponential kernel counterpart with scale $\ell\ell_{XX}$. We then note that in this case, the splitting provides a natural way of systematically providing different scales for different systems, by, for example, making $\ell_{X_iX_j}$ a function of the covalent radii. Additionally, we note other active learning methods using the squared exponential kernel to guide PES exploration \cite{GPMin,koinstinen2017,GarridoPRL2019,denzelGPR2018,AID2020Garrido} could benefit of using kernel \eqref{eq:sqexp} with all $\ell_{X_iX_j}=1$ with no additional retraining since they would obtain similar performance and enforce translation symmetry.

As in previous work \cite{GPMin, AID2020Garrido}, we have used the constant function $m(\mathbf{x})=m_0$ as prior function. We choose to call the diagonal terms in the matrix $\Sigma$ for  $\sigma$ corresponding to the regularization in the forces, and we use $\sigma \ell$ as the regularization of the energies.  

\subsection{Optimization method}

We use the energies and forces from the prediction of the Gaussian
process in equation \eqref{eq:prediction} to guide the searches for the DFT local minimum of the PES. The optimization method we
follow is the one used by GPMin \cite{GPMin} with some variations. 

Starting with the initial atomic configuration, the method computes its DFT energy and forces. This information is used to determine the prior constant $m_0$ and to build a tentative surrogate model of the PES. The method then finds a local minimum of the surrogate PES, computes its DFT energy and forces and includes that point in the training set.
The surrogate model is then updated with the new information, leading to a new location of the minimum, which is subsequently sampled. The iteration terminates when the DFT maximum force on any of the atoms in the system is smaller than a user defined tolerance, as is usual for local optimizers in the ASE package \cite{ase}. The optimization of the surrogate model always takes the structure with the lowest energy in the training set as starting point and then uses the L-BFGS-B optimizer \cite{lbfgsb} as implemented in SciPy \cite{scipy} to find a neighboring local minimum.

In each iteration, the update of the model may include the update of some of its hyperparameters. In the previous section we have introduced the hyperparameters $m_0$, $k_0$, $\sigma$, $\ell$ and $\ell_{X_iX_j}$ for every pair of atomic species $X_i$ and $X_j$ in the atomic structure but not all of them may play an independent role in the prediction of the Gaussian process model \eqref{eq:prediction} and thus not all of them may be updated \cite{GPMin}.

The global scale $\ell$ and the bond scales $\ell_{X_iX_j}$ are not independent of each other, but $\ell$ is rather a global dimensionless prefactor to the bond scales. For this reason, the optimizer regards $\ell$ as a fixed quantity during the optimization of the other hyperparameters.

$k_0$ and $\sigma$ only enter equation \eqref{eq:prediction} in the form of the quotient $\sigma/k_0$, being effectively the same hyperparameter as far as prediction is concerned (note this does not hold for equation \eqref{eq:mll}). 
In addition, we note that the quotient $\sigma /k_0$ is the effective regularization of the Gram matrix $K(X,X)$ and even in the absence of numerical noise in the electronic structure model it needs to be fixed to a small but non-zero value to enable the inversion of the sometimes numerically ill-conditioned Gram matrix, which increases the robustness of the method. In the following section we determine a value of $\sigma/k_0$, which is appropriate for all systems, and the parameter is not updated any further during the optimizations.

Interestingly, the marginal likelihood \eqref{eq:mll} depends on both $k_0$ and $\sigma/k_0$ in a non-trivial way, making it necessary to optimize $k_0$ along with the other hyperparameters to obtain sensible results. In fact, the maximization of the marginal log likelihood \eqref{eq:mll} provides with an analytical expression for the prefactor $k_0$
\begin{equation} \label{eq:k0}
    k_0 = \sqrt{\frac{(Y-\mathbf{m}(X))^T C_{X,k_0=1}^{-1}(Y-\mathbf{m}(X))}{N}},
\end{equation}
if the quotient $\sigma/k_0$ and the scales are kept fixed. A similar expression can be obtained for the prior constant $m_0$:
\begin{equation}\label{eq:m0}
    m_0 = \frac{U^T C_X^{-1} Y}{U^T C_X^{-1} U},
\end{equation}
where $U$ is the prior matrix $\mathbf{m}(X)$ with prior constant $m_0=1$.

In this work we present various flavors of the optimization method  and we compare their performances. The plain version without updates (termed "BondMin" in the following) only differs with default GPMin in the choice of the kernel. It chooses the prior constant to be the maximum of the energies included in the training set and does not update any other hyperparameter. 

We also introduce a method capable of optimizing its own hyperparameters ("BondMin update"). At each step, $k_0$ and $m_0$ are updated using expresions \eqref{eq:k0} and \eqref{eq:m0}. The bond scales, together with $k_0$, are then further updated by numerically maximizing the marginal log likelihood with optimal $m_0$. Here, we follow the strategy used by GPMin in the sense that the values of the hyperparameters are found using SciPy's L-BFGS-B with the constraints of not letting any hyperparameter vary more than 10\% at each step.

A frequently mentioned limitation of Gaussian process regression is the poor scaling of the computational time and memory requirements with the number of points in the training set \cite{williams2006gaussian}. In particular, the use of Cholesky factorization to solve equations \eqref{eq:prediction} and \eqref{eq:mll} results in $O(n^2)$ scaling for the memory and $O(n^3)$ for the computational time (where the scaling factor $n$ is defined as $n=N(3N_\text{atoms}+1)$) in atomic systems training on energies and forces \cite{GPMin,koinstinen2017}. Here we have followed the ideas presented by Garrido Torres \emph{et al.}\cite{AID2020Garrido} as a way to leverage the computational requirements for systems with large numbers of atoms in the unit cell:
\begin{enumerate}
    \item We note that for most molecule-on-surface systems (and more generally, in most systems with a large number of atoms) a significant number of atoms have their positions fixed. Thus, there is no need to train on the forces of the constrained atoms, which can also be masked in the kernel, leading to a scaling factor of $n=N(3N_\text{dyn\_atoms}+1)$, where $N_\text{dyn\_atoms}$ is the number of dynamical atoms.
    \item The problem of predicting the PES for a minimum and its basin with a kernel in the form of equation \eqref{eq:sqexp} mainly depends on the points close to the minimum. In fact, not including distant points may not dramatically decrease accuracy while it may increase the robustness of the method
    \cite{AID2020Garrido, eriksson2019scalable}. This observation allows us to include only the $N_0$ closest points to the current atomic configuration in Euclidean space  in the training set. After the relaxation on the surrogate model has completed, the method checks if there are points that have not been included in the training that are closer than the $N_0$ points used, adds them to the training set, and relaxes the new resulting surrogate model.
\end{enumerate}

All together, the two strategies give a new scaling with $n=N_0(3N_\text{dyn\_atoms} +1)$. This still yields a quadratic scaling for the memory and a cubic one for the computational time, but it is a big improvement in the scaling of the method. Since $N_0$ is now a user-defined fixed number, the computational requirements remain constant instead of growing as the optimization progresses. Additionally, the computational cost remains cubic in time, as for the DFT, but on a smaller variable. 

We have named the method presented in this paper as BondMin when all the sampled points are included in the training set and LBondMin (light memory- BondMin) to the version with the two memory restrictions aforementioned.

\subsection{Computational Details}
We have described the PES using Density Functional Theory (DFT) as implemented in ASE
\cite{ase,ase-paper} and GPAW \cite{gpaw}. All the calculations presented in this work
use RPBE \cite{RPBE} as exchange-correlation functional, a plane wave basis-set, and an
energy cut-off of 600 eV, unless otherwise stated. The Brillouin zone has been sampled with a density of 2.0
k-points per inverse {\AA} in each direction. We have used the projector augmented
wave (PAW) formalism \cite{kresse1999ultrasoft}, using the setup with one valence
electron for sodium and the default one in GPAW otherwise. We impose the convergence
criterium that the maximum change in magnitude of the force over each atom should fall
below $10^{-4}$ eV/{\AA}
to exit the self-consistent field iteration in addition to the default thresholds on the
energy, the density and the Kohn-Sham eigenstates. 

Throughout this paper, we consider a structure relaxed when the maximum force on any atom is below $0.01$ eV/\AA. 

\subsection{Data sets}
The determination of the parameters to the optimizer and the testing of its performance have been done on different sets of atomic structures. The hyperparameters of the method have been determined by training and validating on systems inspired by the training and the test set used to train GPMin, where the elements of group 11 of the periodic table have been substituted by their counterparts in group 10 for the surfaces with adsorbed CO, since CO does not bind to the original Ag and Au surfaces (see the supporting information for more details on this matter). The inclusion of clusters, molecules, bulk structures and surfaces both with and without adsorbates ensures a good overall performance of the method for a large class of systems, preventing overfitting. The method has been tested on two sets of atomic structures containing molecules adsorbed on surfaces.

All the systems considered in this work have been studied in 10 slightly different
initial configurations. The 9 rattled copies of each system are generated by adding white
noise to the atomic positions of the initial one.  The value of the standard deviation of the white noise is specified in the description of each data set.

\subsubsection{Hyperparameter training set}

The hyperparameter training set consists of two different atomic systems: a randomly generated sodium cluster and a CO molecule on a fcc (100) platinum slab. The two original systems have been perturbed with a noise following a Gaussian distribution with standard deviation of 0.1 {\AA}.

\subsubsection{ASE/GPAW test set}

The ASE/GPAW test set is the same as the test set presented in reference \cite{GPMin} and which is available in the GPAW webpage \cite{gpaw-opt-test}, with the exception that silver has been substituted by palladium in the CO on a surface test. The set consists of two molecules, hydrogen molecule and pentane molecule, the bulk structure of copper fcc in a $2\times 2 \times 2$ supercell, shaken; a two-layer distorted copper fcc (111) slab, a 13 atom aluminum cluster and two adsorbates: a carbon atom on a $2\times 2\times 2$ fcc (100) copper slab and the aforementioned CO molecule on a $2 \times 2 \times 2$ fcc (111) palladium slab.

\subsubsection{MS5: Small molecules on surfaces data set}

This data set is made up of five adsorbates with up to 5 atoms at three different adsorption locations.
It consists of two molecules, water and nitrogen dioxide
on palladium fcc (100) surfaces and hydroxyl, hydroxymethyl and
methyl radicals on copper fcc (100) surfaces. We have used
$2\times 2\times 2$ slabs to represent the surfaces and constrained the movement of the atoms in the bottom layer.
For each system, we have studied three initial highly symmetric bonding sites of the molecule or radical, those termed ``on top", ``hollow" and ``bridge", as implemented in ASE.
The G2 set has been used to obtain the initial structures of the molecules \cite{G2}.
Each of the original systems has been perturbed with a noise following a Gaussian distribution with standard deviation of 0.07 {\AA}.

\subsubsection{C3-4S: 3 and 4 carbon organic molecules on surfaces data set}

The C3-4S set contains two different molecules: acrylic acid \cite{acrylic_acid} (CH${}_2=$CHCOOH) molecule on a fcc (111) $4\times 3$ palladium surface in the ``on top" position and butanethiolate \cite{butaethiol} radical (C${}_4$H${}_9$S${}^*$) on a fcc (111) $3 \times 3$ gold surface in the ``hollow" position. Both surfaces are modelled by a 3 layer slab, with the atoms in the bottom layer being kept fixed during the relaxation.
Thus, the  acrylic acid system has 33 dynamical atoms (45 in total) and the butanethiolate one, 32 (41 in total).
We have solved the electronic structure problem with increased convergence for these two systems: in addition to raising the plane wave energy cut-off to 800 eV, we have added the additional threshold for the termination to the self consistent field iteration that the change in the energy in the last 3 iterations should be less than $10^{-6}$ eV per valence electron.

Each of the original systems has been perturbed with a noise following a Gaussian distribution with standard deviation of 0.07 {\AA}.

\subsection{Selection of the hyperparameters}
 
In this section we present the values of those hyperparameters that should remain fixed during the optimization as well as the initial values of the remaining ones. The selection of these values has been done in a two step process: First, we investigate the performance of the relaxation method on the hyperparameter training set with different sets of hyperparameters. In a second step, we have chosen the values of the hyperparameters whose performance is more consistent across the more diverse ASE/GPAW test set among those that performed the best on the hyperparameter training set. Thus, the ASE/GPAW test set acts as a validation set here, preventing overfitting and ensuring increased robustness of the method.

For the method with fixed hyperparameters, we have modelled the bond scales as the average of the covalent radius $r^c$ as tabulated by Cordero \emph{et al.} \cite{Cordero} of the species:
\begin{equation}\label{eq:model}
\ell_{X_iX_j}=\frac{r^c_{X_i}+r^c_{X_j}}{2},
\end{equation}
but the method allows for a user-defined model. In particular, we have further tested the product of covalent radii 
$\ell_{X_iX_j}=\sqrt{r^c_{X_i}r^c_{X_j}}$, which, under the appropriate 
choice of the rest of the hyperparameters, did not produce a qualitative 
improvement when it was tested on the validation set.

\begin{figure}[t]
    \centering
    \includegraphics{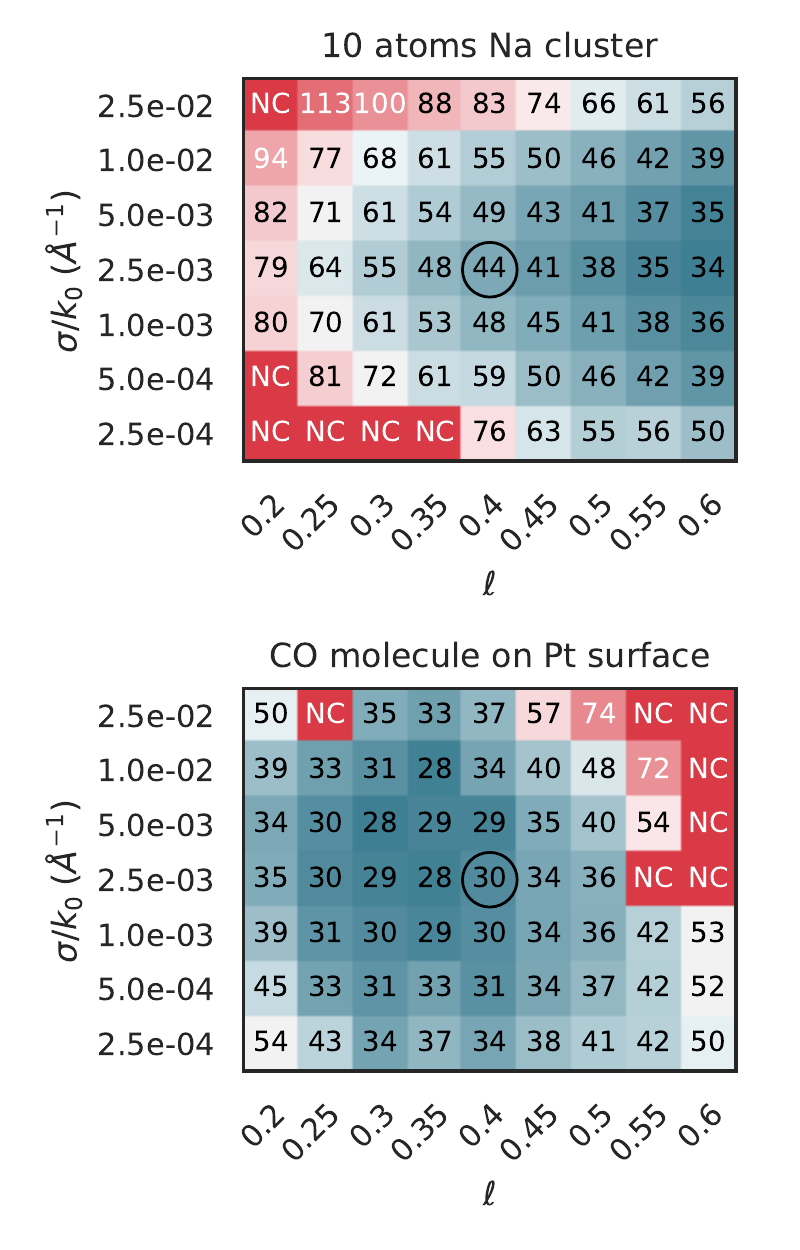}
    \caption{Average number of DFT calculations required to relax the structures in the training set for the method with fixed bond scales as a function of the effective regularization $\sigma/k_0$ and the global dimensionless scale $\ell$. The label "NC" stands for at least one relaxation failed with that sets of hyperparameters. The circle marks the values of the hyperparameters that have been used in the subsequent calculations in this paper with fixed bond scales. Both the colors and the figure in the heat map show the average number of steps.}
    \label{fig:train}
\end{figure}

The values of the other hyperparameters $\sigma/k_0$ and $\ell$ have been chosen such that they minimize the average number of DFT calculations necessary to relax the structures in the training set. The results are shown in Figure \ref{fig:train}.

Even with the anisotropy introduced with the inclusion of a model for the bonds in the Gaussian process regression, Figure \ref{fig:train} shows the metallic cluster still prefers a longer value of the global scale, $\ell$; while for the molecule on the surface it is more favorable to choose a shorter value. Both systems prefer a value for the regularization $\sigma/k_0 \sim 10^{-3} \text{\AA}^{-1}$. Expecting a value for the prefactor of the order $k_0\sim 1\text{eV}$ this leads to $\sigma \sim 10^{-3} \text{eV}/\text{\AA}$. This is a reasonable value, since an order of magnitude higher would conflict with the convergence threshold of the optimizer ($0.01\text{eV}/\text{\AA}$) and the forces are converged with precision $10^{-4} \text{eV}/\text{\AA}$.
We have chosen the values $\ell=0.4$, $\sigma=2.5\cdot 10^{-3} \text{eV}/ \text{\AA}$ and $k_0=1\;\text{eV}$.

We have also used the average of the covalent radii in equation \eqref{eq:model} as the initial value for the bond scales when the model is set to update them (and tested that the square root of the product does not produce a significant improvement on the validation set when the other hyperparameters are trained accordingly). For the initial value of the prefactor of the kernel, we have chosen the value that was found for GPMin during training, $k_0=2\; \text{eV}$.

\begin{figure}
    \centering
    \includegraphics{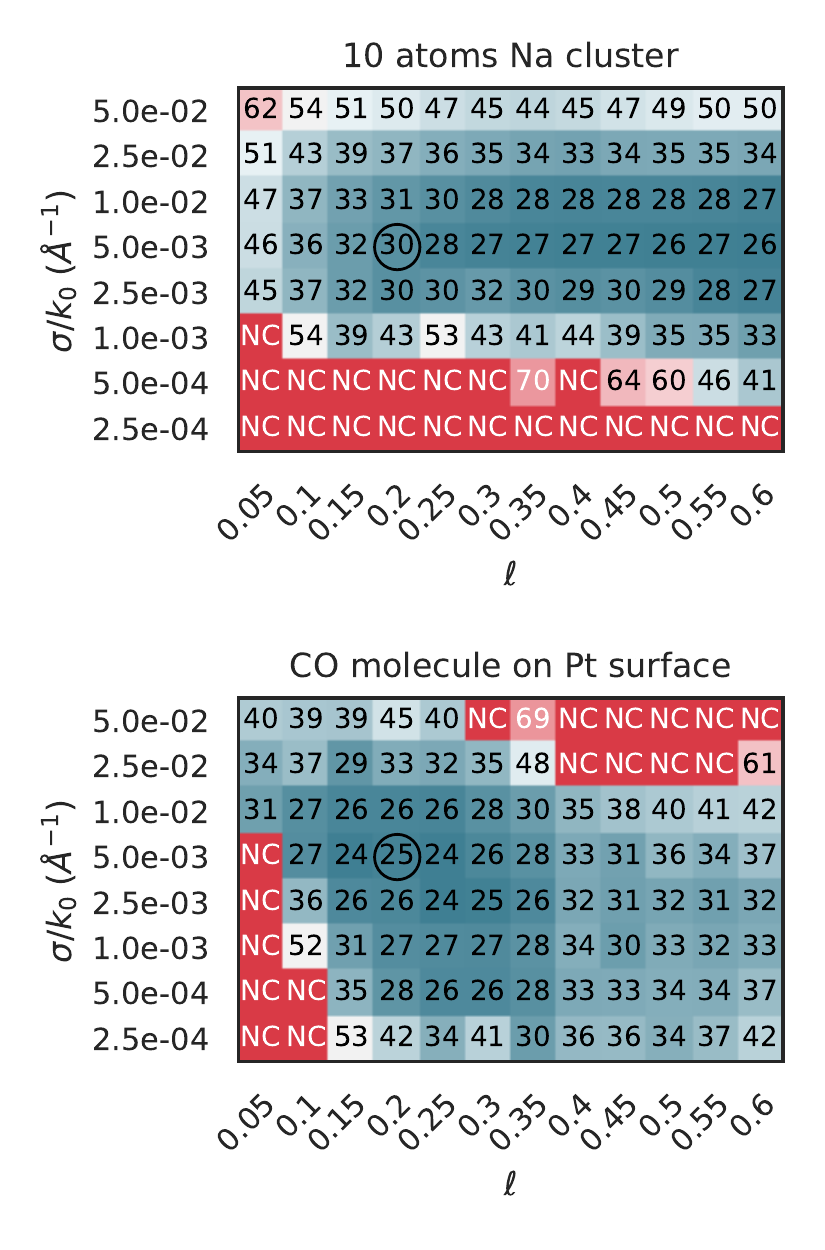}
    \caption{Average number of DFT calculations required to relax the structures in the training set for the method with updated bond scales as a function of the effective regularization $\sigma/k_0$ and the global dimensionless scale $\ell$. The label "NC" stands for at least one relaxation failed with that sets of hyperparameters. The circle marks the values of the hyperparameters that have been used in the subsequent calculations in this paper with updated bond scales. Both the colors and the figure in the heat map show the average number of steps.}
    \label{fig:train_update}
\end{figure}

As for the version of the method with updated bond scales, the values of $\ell$ and $\sigma/k_0$, which remain fixed, have been determined by analyzing the performance of the method over the training set, as shown in Figure \ref{fig:train_update}.

Figure \ref{fig:train_update} shows that it is easier to find the optimal performance by maximizing the marginal log-likelihood if the initial scales in the GPR  underestimate their value, as compared to overestimates, which had already been observed in previous work \cite{GPMin}. This results in an almost flat number of steps as a function of $\ell$ for the sodium cluster, which prefers overall long scales; but a sharp minimum for the CO on platinum. Our investigations show that the values $\sigma = 2.5 \cdot 10
^{-3} \text{eV}/\text{\AA}$ and $\ell=0.2$ result in a good overall performance in both the training and the validation sets.

\begin{figure*}
    \centering
    \includegraphics[width=0.9\textwidth]{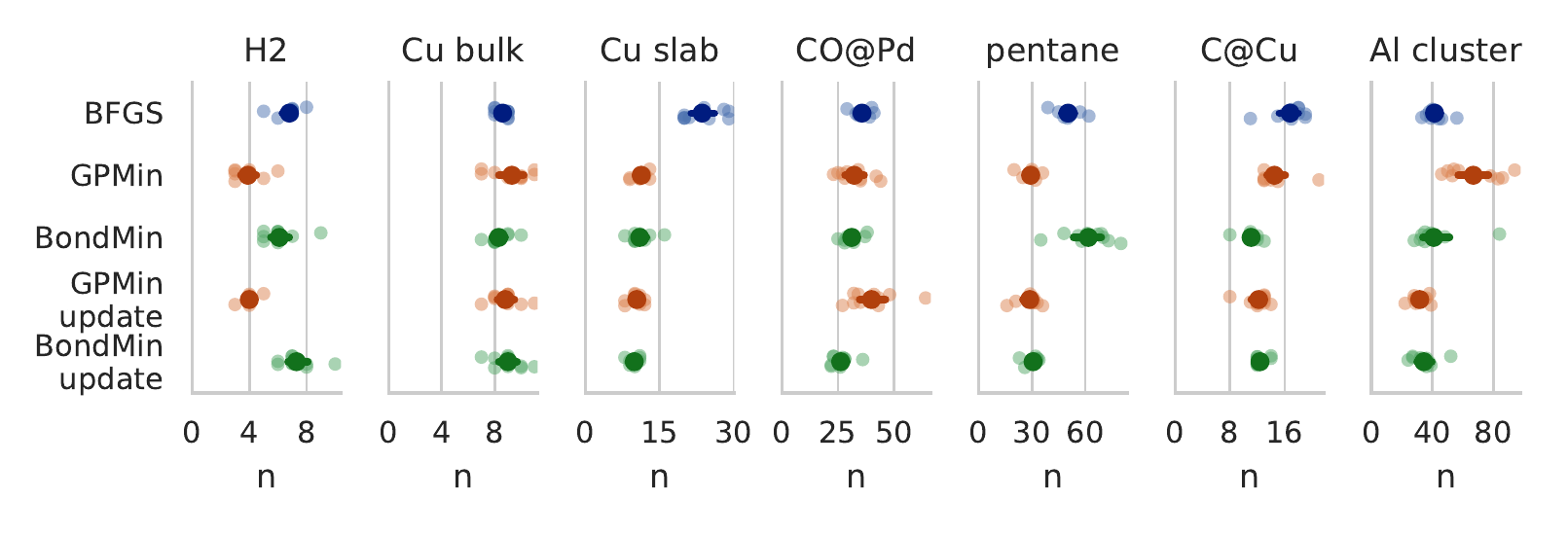}
    \caption{Number of steps needed to relax the systems in the validation set for the different optimizers. Lighter markers represent individual runs, while the darker ones mark the average number of runs and its error bar.}
    \label{fig:validation}
\end{figure*}

We conclude the methods section by commenting on the performance of the method with and without hyperparameter updates on the ASE/GPAW test set. The performance of each method, along with  BFGS with Line Search as implemented in ASE and both the default version of GPMin and GPMin with hyperparameter updates are shown in Figure \ref{fig:validation}. We compare with BFGS with line search because in a previous study this was clearly the best ASE or SciPy optimizer on this test set \cite{GPMin}.

From Figure \ref{fig:validation} we observe that the performance of the BondMin method, especially with parameter updates, is in general similar to that of GPMin for this class of systems. As noted in the Gaussian process regression section, BondMin reduces to GPMin with a different set of hyperparameters for unary systems.
We believe this to be the cause for the bad performance on the H${}_2$ system: 
a scale of 0.12 $\text{\AA}$ for the method without updates and an initial scale of 0.6 $\text{\AA}$ 
for the method with updates force the method to take (at least, initially) very short steps in
configuration space, which makes it difficult to compete with 4 steps in average GPMin. 
However, we attribute the increased performance in the aluminum cluster to the same effect, where we believe the new corrected global scale is initially close to optimal.

In contrast, we note the worsening of performance for the pentane molecule, but we also note it is improved when hyperparameters are allowed to update. In addition, it seems the new kernel improves the results on molecules on surfaces the most; especially CO on palladium, which seems to be a difficult problem for GPMin.  

We see, that the performance of the updated BondMin is rather similar to the updated GPmin. In the cases with only one type of interatomic bonds the two methods should behave similarly. However, in the case of CO/Pd, where both metallic, molecular, and molecule-metal bonds are present BondMin seems to be superior.

\section{Results}

\begin{figure*}[t]
    \centering
    \includegraphics[width=\textwidth]{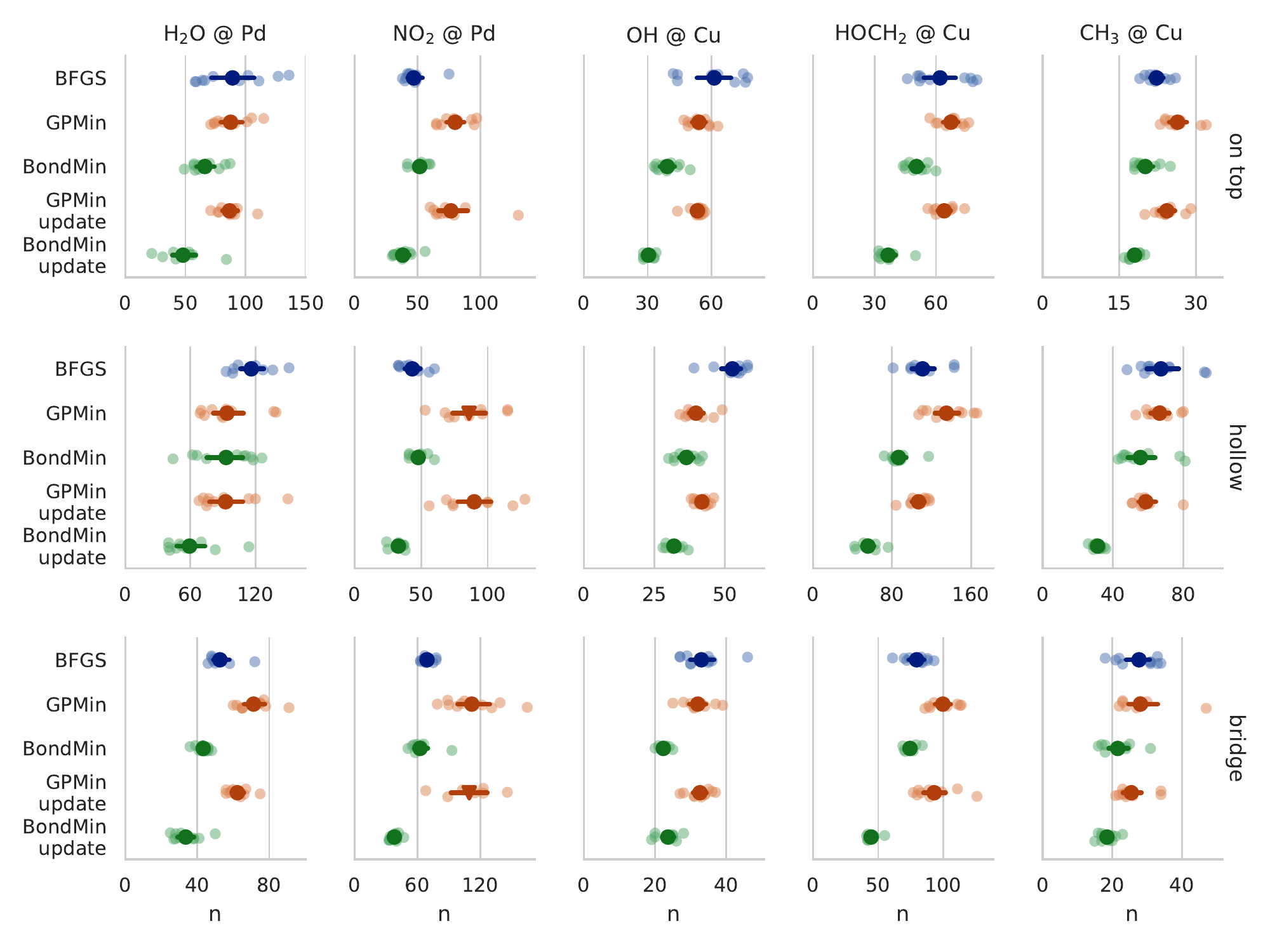}
    \caption{Number of steps needed to relax different molecules and radicals on fcc (100) slabs for the different optimizers. Lighter markers represent individual runs, while the darker ones mark the average number of runs and its error bar. All ten runs converged for systems marked with $\bullet$, while the averages for systems where at least one relaxation failed are marked with $\blacktriangledown$.}
    \label{fig:results_g2}
\end{figure*}

In the tests on the validation set above, the BondMin optimizer shows an improved performance on a particular subclass of systems: molecules on surfaces. To illustrate this further, we have  studied the performance of BondMin with and without bond scales updates as compared to other optimizers for data sets MS5 and C3-4S involving molecules and radicals on surfaces. 
The results on MS5 are shown in Figure \ref{fig:results_g2}. 

We note that the optimizers of the GPMin family do not show a consistent improvement on this test set as compared to BFGS. In particular, GPmin shows a relatively poor performance on the structures in the bridge initial positions and the NO${}_2$ molecule for all initial positions. In fact, in three of the runs for GPMin with hyperparameter updates on NO$_{2}$@Pd with starting bridge position the relaxation was terminated and marked as failed after more than 210 steps had been taken without finding the minimum.  An additional NO${}_2$@Pd relaxation has failed with the GPMin optimizer and hollow initial position, in this case because the model was not able to predict a low energy configuration in 30 consecutive steps. These failed optimizations are marked with an inverted triangle in Figure \ref{fig:results_g2}.

In contrast, the BondMin family of optimizers seems to perform well in this test set. The BondMin version without hyperparameter updates consistently shows similar or lower number of steps than BFGS, and it is also competitive compared to GPMin with hyperparameter updates in number of steps, but with a smaller computational cost.  

The BondMin with hyperparameter updates optimizer exhibits the lowest number of steps needed to relax all the systems in this test set. As compared with BFGS, it shows an average reduction of over 40\% on the number of steps necessary to relax a molecule on a surface. The relative reduction in the number of steps seems to be more pronounced on those systems where the number of steps required by BFGS is large, reaching a factor of 2 reduction for  H${}_2$O and HOCH${}_2$ on the hollow initial and OH radical in the on top position, and a factor 2.15 reduction for the CH${}_3$ radical in the "hollow" position.

Furthermore, the BondMin optimizers also show a reduction of the spread in the number of steps among the 10 slightly rattled initial conditions as compared to the other methods. BondMin with hyperparameter updates also shows an average reduction of the standard deviation of the number of steps of over 40\% as compared with BFGS. Even though the standard deviation is comparable with the one of BFGS for the water molecule, we note that for some of the systems it can reach up to a factor 5 reduction, with standard deviations of only 2 or 3 steps in a large fraction of the tests.

We now turn to the results of the limited memory approach (LBondMin). The performance of LBondMin has been studied on the hyperparameter training set systems (where the optimal hyperparameters are known) as a function of the training set size and further tested on two large systems as illustrations.

\begin{figure}[t]
    \centering
    \includegraphics{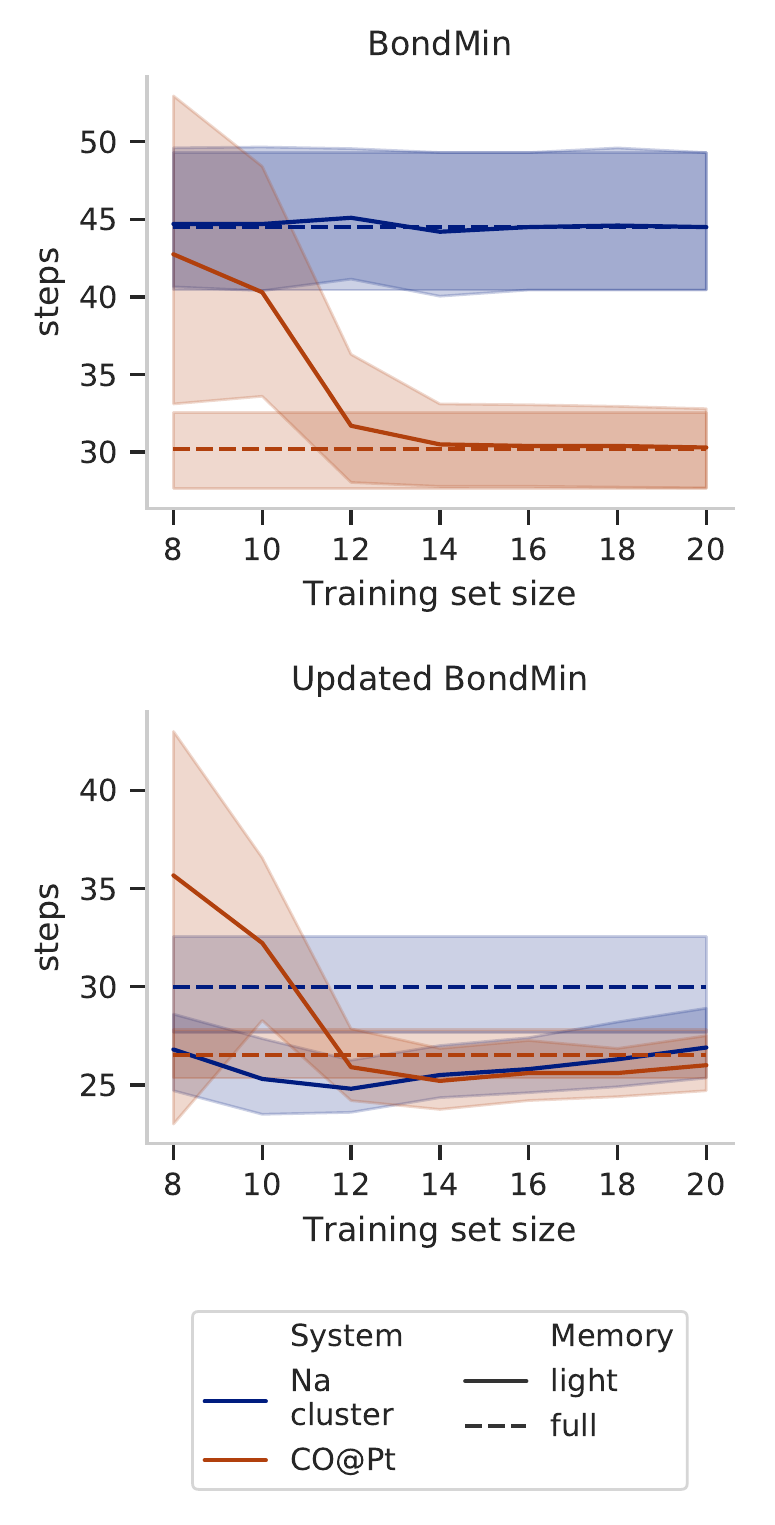}
    \caption{Average number of steps of the light memory BondMin optimizer as a function of the size of the training set for the two training systems. The shaded areas indicate the variance on the mean distribution as estimated from ten different runs. The dashed line shows the results of BondMin with no memory restrictions for the same systems.}
    \label{fig:train_mem}
\end{figure}

The results for the hyperparameter training set (a sodium cluster and the CO on Pt) are shown in
Figure \ref{fig:train_mem}. The training set size for the PES is now limited to the last $N_0$ configurations in the light memory version, and the figure shows the number of required minimization steps for different values of $N_0$. The full memory version is included for comparison.

In the relaxations without hyperparameter updates the performance seems to saturate to the value of the full memory method relatively fast. 
In particular, the performance for the sodium cluster is qualitatively indistinguishable between the full memory and low memory versions for the range of
training set sizes studied. We note that the performance of the full memory method could have been
achieved with a fourth of the training images for the sodium cluster  and with half as many training images for the CO on Pt. 

Including the update of the hyperparameters, a different picture emerges: the reduction of the number of points in the training set can lead to a reduction of the number of steps needed to relax the structure in the systems studied. This is particularly significant for the sodium cluster, where the average number of steps is reduced by 10-17\% with respect to the full memory method in all of the investigated range of training set sizes. Again, for both systems one could have used about half of the number of points in the training set, with a modest boost in performance as an effect.

\begin{figure}[t]
    \centering
    \includegraphics{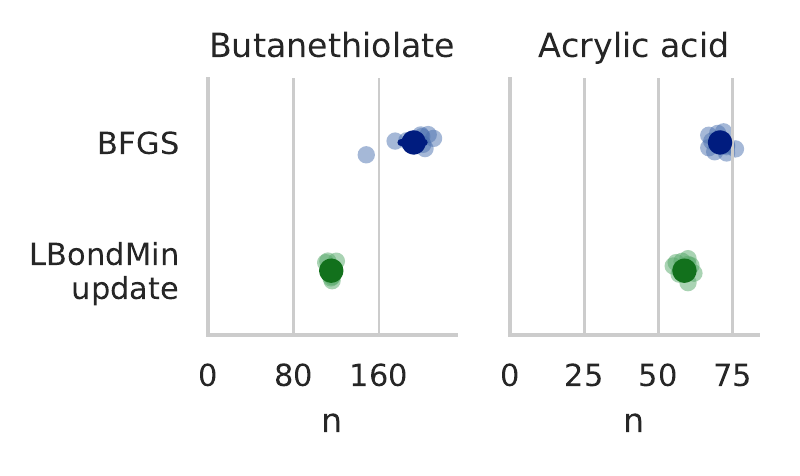}
    \caption{Number of steps needed for different optimizers to relax the molecules on surfaces in data set C3-4S. Lighter markers represent individual runs, while the darker ones mark  the  average  number  of  runs  and  its  error  bar.  }
    \label{fig:mem_large}
\end{figure}

We illustrate the application of the light memory approach on the two systems of larger size in the C3-4S data set.  The results of the optimization can be found in Figure \ref{fig:mem_large}. 

For these systems, we have limited the size of the training set to 20 points. The results obtained are consistent with those shown in Figures \ref{fig:results_g2} and \ref{fig:train_mem}: The BondMin optimizer results in a significant reduction of the number of steps needed to relax the system as compared to BFGS even by reducing the number of points in the training set by a factor 5 (in the case of butanethiolate radical). Thus, it shows the reduction of the training set size results in a reduction of computational cost while retaining the performance.

\section{Discussion}

The BondMin optimizer, especially with hyperparameter updates, shows superior performance to the GPMin and BFGS line search optimizers for molecules on surfaces, and at the same time comparable performance on a broader range of atomic systems. In particular, it shows speed-ups of up to more than a factor of two on the moderate size adsorbates as compared with BFGS. Moreover, the speed-up seems to increase in those systems where the performance of BFGS is poor, and shows robust behavior with small differences in performance for different initial configurations. Thus, the performance of the BondMin method over different systems and initial conditions is not only superior, but more consistent and reliable, as compared with the other methods presented in this article.

We ascribe these improvements to two factors. Firstly, we believe that the ability of the Gaussian process regression to capture both harmonic and anharmonic regimes improves the description of the PES close to saddle points reducing the number of sample points needed to get out of them. This characteristic is shared with other GPR methods such as GPMin, as a contrast to the quadratic model in BFGS.

\begin{figure}[t]
    \centering
    
    

    
    
    
    
    \includegraphics{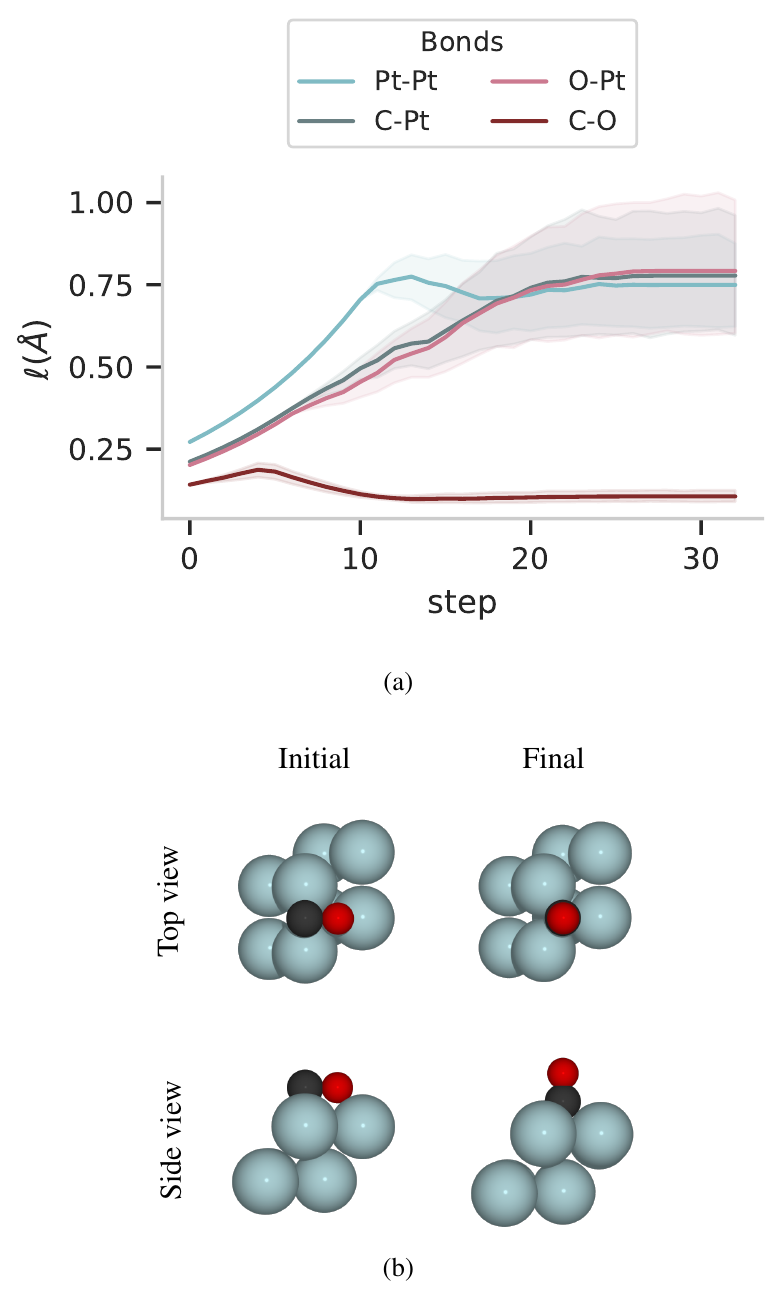}
    \caption{CO on platinum during the optimization. (a) Evolution of the bond scales as a function of optimization step. The hyperparameters used are those chosen in Figure \ref{fig:train_update}. The solid line represents the average over 10 runs and the shaded area, the uncertainty over the average.(b) Atomic structures of the initial and final atomic configurations. Only the atoms in the unit cell are shown.}
    \label{fig:evolution}
\end{figure}

\begin{figure}[t]
    \centering
    \includegraphics{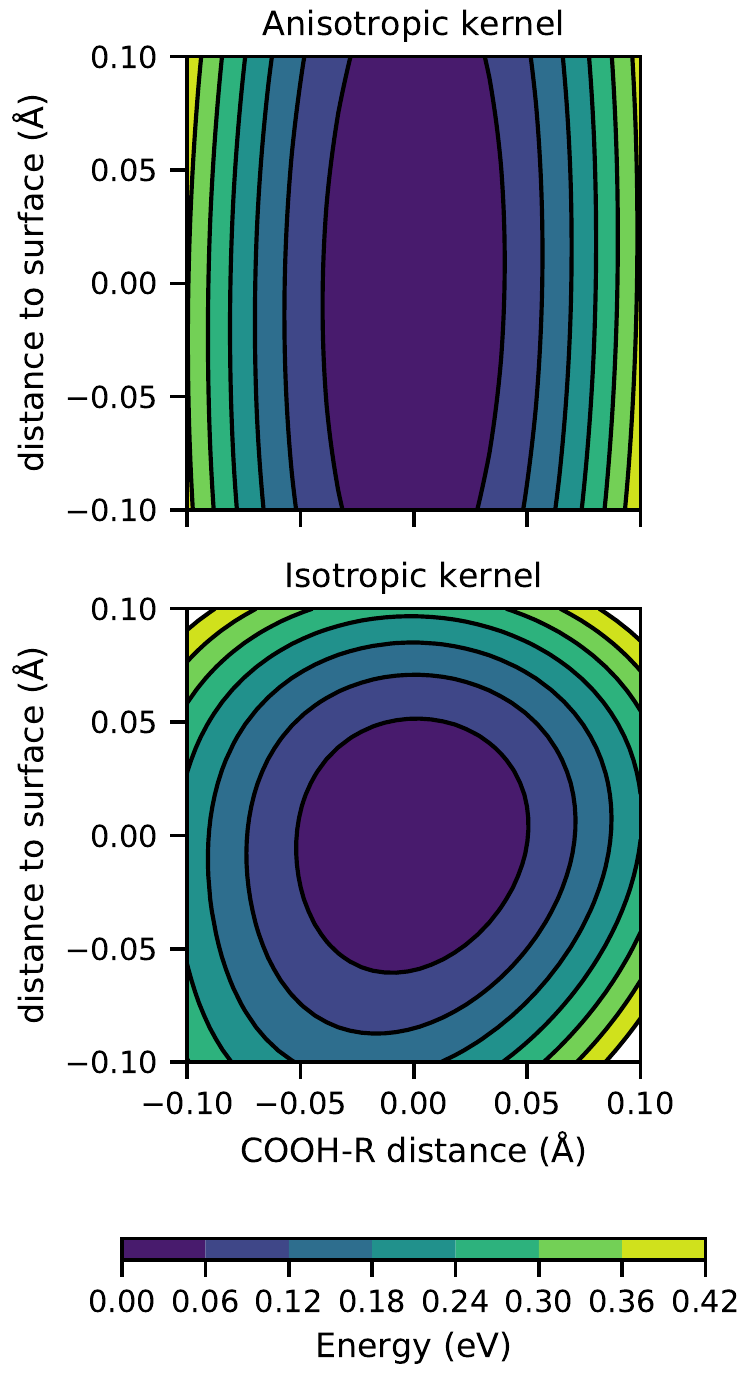}
    \caption{Surrogate models of the potential energy surface of acrylic acid on palladium around the minimum for two different kernels. The x axis represents the dissociation of the molecule  into the carboxyl and the vinyl radicals, while the y axis accounts for the variation of the adsorption distance between molecule and the surface. The anisotropic kernel in the top panel corresponds to the one used by BondMin while the isotropic is the one used by GPMin.}
    \label{fig:contour}
\end{figure}

Secondly, BondMin is able to adapt to anisotropic potential energy surface landscapes with fewer points, as compared with isotropic kernels, which would need a large number of points to describe an anisotropic landscape. We believe this capacity is the key to success for problems involving molecules on surfaces, since they typically involve a combination of stiff and soft bonds that are difficult to capture for GPMin. We also suggest that the reduced number of parameters needed to model the anisotropy as compared to BFGS (i.e. a few bond scales vs. the full Hessian) may contribute to the improved performance of BondMin as compared with BFGS line search. We further illustrate this point in Figures \ref{fig:evolution} and \ref{fig:contour}.

The initial bond scales obtained from the covalent radii of the atoms in the system provide a reasonable preconditioning for the
different scales involved in the problem, as illustrated in Figure \ref{fig:results_g2} for the BondMin optimizer without updates of the hyperparameters. However, the ability to update the hyperparameters results in a better model of the potential energy surface. Figure \ref{fig:evolution} (a) shows the evolution of the scales $\ell\cdot\ell_{X_iX_J}$ for the different bonds in CO on platinum, with the hyperparameters found during training. The scales start at 0.2 times the average of the covalent radii of the two species and evolve a maximum of a 10\% every step to maximize the log marginal likelihood. As evidenced in Figure \ref{fig:evolution} (b) the main task of the optimizer is to rigidly rotate the CO molecule from being parallel to perpendicular to the surface of the slab.

The carbon-oxygen scale starts at 0.14 $\text{\AA}$ and it ends at 0.11 $\text{\AA}$, on average. These are low values compared to the other bonds in Figure \ref{fig:evolution} (a). The CO scale also presents comparably low variation between the initial and final scales and low spread, compared to the other bonds. We attribute this to the fact that there is only one CO bond in the atomic structure and to the stiffness of the CO bond.
We conclude from this that the machine learning algorithm is able to correctly learn to separate the stiff covalent bonds in molecules (i.e. those exhibiting relatively fast variations in energy as the bond distance is changed) from softer molecule-surface and metallic bonds.

The other bonds, O-Pt, C-Pt  and Pt-Pt; start at 0.20, 0.22 and 0.27 $\text{\AA}$, respectively. These seem to be underestimates, since by the end of the run the average value over 10 runs ends up being around 0.77 $\text{\AA}$ for these three bond scales. Initially, the  Pt-Pt bond scale grows faster than the O-Pt and C-Pt scales, reaching the final value earlier. We attribute this behaviour to the fact that the Pt slab is initially very close to the relaxed configuration and it can thus relax to it very fast, while the rotation of the molecule provides an increasingly diverse training set as the optimizer gathers more data. It can be noted that these scales are shorter than those found by GPMin on the sodium cluster, which were always longer than 1 $\text{\AA}$, which might have to do with the relatively soft bonds in the cluster.

The optimized molecule-Pt scales show a larger variation across the 10 different relaxations than the Pt-Pt ones by the end of the run. We attribute this to the ability of the method to adapt itself to fit the minimization trajectory the best: differences in the configurations sampled by the optimizer might lead to differences in the relevance of different "bonds" in different training structures and thus leading to slightly different models. 

The introduction of different scales associated with the different bonds leads to models which can capture anisotropy better with fewer data points, as illustrated in
Figure \ref{fig:contour} for acrylic acid on palladium (data set C3-4S). 
The figure shows the level sets of the models of the PES underlying 
BondMin and GPMin, respectively, around the minimum.  
Both regressions have been trained on the data from the trajectory of one
of the LBondMin energy minimizations for this system, and the 
hyperparameters have been fully optimized. The plot shows the variation 
in the potential energy when breaking the single C-C bond which splits the 
molecule into the carboxyl and vinyl radicals and translating the whole molecule along the z direction, away or towards the surface.

The scales for the bonds between the C, O and H atoms and palladium saturate at between 3.6-3.8 $\text{\AA}$, in contrast with the 0.14 $\text{\AA}$ of the C-C bond. This produces weights in equation \eqref{eq:dist} between 600 and 700 times smaller for the translation of the molecule with respect to the surface as compared with its dissociation, resulting in a more anisotropic PES, as illustrated in the top panel of Figure \ref{fig:contour}. In contrast, the optimization of the scale in GPMin over these same training points yields a value of 0.6$\text{\AA}$, which is a compromise between the long scales and short scales found by BondMin. The GPMin model for the potential energy surface does not capture the different nature of the two bonds with this amount of information, even when the training set contains the minimum and neighboring points. The Gaussian process model underlying GPMin is less reliable at extrapolating and, as a result, the GPMin optimizer would in general need more points to describe the surface and find its minimum.

The surrogate model of the potential energy surface presented in this work is invariant under rigid translations of the whole system, which turns out to be both a blessing and a curse. Even though the underlying physics is translationally invariant, we note that the numerical solution to the Kohn-Sham equations does not need to fully obey this. For instance, if the grid in real space is too coarse, this might lead to and egg-box effect \cite{nogueira2003tutorial}, resulting in a small translation-dependent spurious potential.

As a result, combining a DFT method whose parameters are not finely tuned together with a very tight threshold for the optimization step can result in the optimizer failing to converge. To our experience, this feature becomes particularly relevant for systems where some of the atoms are constrained to stay fixed: the local minimum might be in a direction that would represent a translation, and the surrogate model would never be able to capture that.

We suggest that, in most cases, the solution to this situation is to reconsider the accuracy that is needed for the particular application and either increase the convergence of the DFT method or relax the tolerance for the convergence of the optimization method. Notwithstanding, we have considered some possible alternative solutions when the above is not possible.

One possibility would be to explicitly break the translational invariance of the model, by adding a soft-mode for rigid translations. This can be done by defining a new matrix $\tilde{G}$:
\begin{equation}
\tilde{G} = G + \ell_T^{-2} (\mathbf{t}_x\mathbf{t}_x^T + \mathbf{t}_y\mathbf{t}_y^T + \mathbf{t}_z\mathbf{t}_z^T),
\end{equation}
where $\ell_T$ is the scale for the translation mode and $\mathbf{t}_x$, $\mathbf{t}_y$ and $\mathbf{t}_z$ are the unitary vectors generating the translations of the system along the three axis.

Another possibility is to redefine convergence: by defining the corrected
forces on the atoms to be the DFT forces minus the average force over all
atoms, such that the sum of all forces is zero, one can redefine the
convergence criterium of the optimizer as having the maximum corrected
force among the atoms to fall under a certain threshold.
This approach results in an approximate best structure given the circumstances,
which could then be finely tuned with a more precise DFT
method. We note that when using this approach, the Gaussian process still needs to be trained on the uncorrected forces (those not being
translationally invariant) since training on corrected forces would introduce an energy-force noise term in the model.

We now turn to a discussion of the approach where the number of training points is limited to a constant number. As far as we know, there is no easy rule of thumb to determine the number of neighboring points one should include in the training in order to obtain the optimal speedup. The sodium cluster and the CO molecule on platinum in Figure \ref{fig:train_mem} have the same number of atoms in the system and show different optimal numbers of points in the training set. Moreover, the potential gain (if any) compared to using the full data set also seems to vary from system to system.

Considering the poor scaling of the full memory approach with the number of atoms in the system, we consider the better scaling of the light memory approach at no significant reduction of performance for a wide range of values of the training set size makes it the method of choice for large systems. In such cases, the number of points to include in the training set should be chosen by the user under a consideration of the computational resources available for the problem at hand.

Let us finally note, that all the molecules on surfaces discussed in this article bind to the surfaces using RPBE for the exchange-correlation energy. We have chosen not to show systems that do not bind since, for such a system, the PES does not have a clear and well defined minimum. This makes step-counts as a measure of performance difficult to interpret. We have included some tests on non-bonding systems in the supporting information, where we show that BondMin still performs well in finding the minimum for such systems.

\section{Conclusion}

We have presented three versions of a local optimization method based on a model composed of preconditioned radial functions: a full memory version with hyperparameter updates, the same method without updates, and a light version with less memory requirement. We have shown that the full memory version with hyperparameter updates reduces the average number of steps needed to relax molecules on surfaces in a robust manner, with potential speed-ups of up to a factor 2 compared to BFGS, depending on the system. The light memory version works with a reduced training set, which might be necessary for large systems in the present implementation. Surprisingly, the limitation of the training set might in some cases lead to superior performance.
A reimplementation of the method using parallelization and distributed techniques \cite{NIPS2018_gpytorch,NIPS2019_million},  would make the method benefit from the kind of speed-ups most DFT implementations are already taking advantage of when executed at large supercomputing facilities.

The method presented obtains comparable improvements over standard optimizers to those presented in other works using Gaussian Processes \cite{denzelGPR2018,Hauser2020Internal} on other classes of materials, while retaining a good overall performance on a wide class of systems. In several references, the boost in performance has been attributed to the use of non stationary kernels \cite{Hauser2020Internal,koistinen2019nudged} as a way to include relevant information relative to chemical bonding. In contrast, we show that the preconditioning of an isotropic stationary kernel can in fact include a crucial fraction of the bond information in an unbiased way. This is particularly useful for adsorption systems, where the bonds inside the molecule, inside the substrate, and between the two might involve different length scales. As proposed by Meyer and Hauser \cite{Hauser2020Internal}, the combination of preconditioning with non-stationary kernels might bring a further reduction of the computational time needed in local explorations of the potential energy surface.

As discussed by Garrido Torres \emph{et al.} \cite{AID2020Garrido}, a further gain compared to traditional methods can be achieved by using the method for several calculations on the same system, as for example, when relaxing the same molecule on different sites. The first relaxation would exhibit a speed-up comparable to the one discussed in this paper, and the subsequent ones would benefit of better initial estimates of the hyperparameters as well as of the energies and forces from previous relaxations.

Finally, we note that the choice of the initial preconditioning as the average of the covalent radii is physically reasonable but also somewhat arbitrary. We have shown this choice improves the performance in systems where the PES is anisotropic and the number of steps is large, but we have also reported that it severely underestimates the ratio between molecular bond scales and the scales of molecule-surface bonds. Along this line we note that the implementation of the method is flexible enough to allow for other user-defined choices of the initial scales, for example van der Waals radii \cite{alvarez2013cartography}, results from previous similar calculations, or parameters extracted from semi-empirical models \cite{tadmor2011potential}. 

\section*{Supplementary Material}
See supplementary material for the binding energies and number of steps in the relaxations with different
optimizers for additional systems.


\begin{acknowledgments}
We acknowledge support from the VILLUM Center for
Science of Sustainable Fuels and Chemicals, which is funded by
the VILLUM Fonden research grant (9455). We also acknowledge support from the U.S. Department of Energy, Chemical Sciences, Geosciences, and Biosciences (CSGB) Division of the Office of Basic Energy Sciences, via Grant DE-AC02-76SF00515 to the SUNCAT Center for Interface Science and Catalysis.
\end{acknowledgments}

\bibliography{main}

\end{document}